\documentclass[runningheads,a4paper]{llncs}
 
\usepackage{microtype}
\usepackage{todonotes}
\usepackage{amsmath}
\usepackage{amsmath,amsfonts,amssymb} 
\usepackage{url}
\usepackage{stmaryrd}
\usepackage{hyperref}
\usepackage{enumerate}
\usepackage{todonotes}
\usepackage{wrapfig}
\usepackage{pgfplots}
\usepackage{multirow}
\usepackage{verbatim}

\usepackage{amssymb,amsmath}
\usepackage{enumerate}
\usepackage{paralist}
\usepackage{extarrows}
\usepackage{mathtools}

\usepackage{algorithm}
\usepackage{algorithmic}
\usepackage{listings}
\usepackage[title]{appendix}
\usepackage{marginnote}
\usepackage{listings}
\usepackage{url}

\newcommand{\M}{\mathbin{\mathbf{M}}} 
\def\Masking{\preceq_{m}}
\def\WeakMasking{\preceq^w_{m}}
\def\FMask{f_{m}}
\def\DeltaMask{\delta_{m}}

\def\Refuter{R}
\def\Verifier{V}
\def\ErrorSt{s_{err}}
\def\Sup{\sup}
\def\Inf{\inf}

\def\pr#1#2{\mbox{pr}_{#1}(#2)}
\def\SigmaF{\Sigma_{\mathcal{F}}}
\def\SigmaM{\Sigma_{\mathcal{M}}}
\def\MaskD{\textsf{MaskD}}
\def\val{\mathop{\textup{val}}}

\renewenvironment{proof}{\noindent\textbf{Proof.}}{\hfill$\diamond$}

\usepackage{color}

\definecolor{lightblue}{RGB}{231,255,255}
\definecolor{lightred}{RGB}{255,224,224}
\definecolor{lightgreen}{RGB}{224,255,224}
\definecolor{lightyellow}{RGB}{255,255,224}
\definecolor{lightpurple}{RGB}{255,224,255}
\definecolor{darkerred}{RGB}{64,0,0}
\definecolor{darkred}{RGB}{128,0,0}
\definecolor{darkblue}{RGB}{0,0,128}
\definecolor{darkgreen}{RGB}{0,128,0}
\definecolor{darkpurple}{RGB}{128,0,128}
\definecolor{black}{RGB}{0,0,0}

\makeatletter
\def\THICKhrulefill{\leavevmode \leaders \hrule height 5pt\hfill \kern \z@}
\makeatother

\urldef{\mailsa}\path|{dargenio,rdemasi}@famaf.unc.edu.ar|
\urldef{\mailsb}\path|{pcastro,lputruele}@dc.exa.unrc.edu.ar|

\graphicspath{{./figures/}}

\begin{document}

\title{Measuring Masking Fault-Tolerance%
  \thanks{This work was supported by grants ANPCyT PICT-2017-3894
    (RAFTSys), SeCyT-UNC 33620180100354CB (ARES), and the ERC Advanced
    Grant 695614 (POWVER).}}

\titlerunning{Measuring Masking Fault-Tolerance} 

\author{Pablo F. Castro \inst{1,3} \and
Pedro R. D'Argenio \inst{2,3,4} \and 
Ramiro Demasi \inst{2,3} \and 
Luciano Putruele \inst{1,3} }

\authorrunning{P.F. Castro et al.}
\institute{Departamento de Computaci\'on, FCEFQyN, Universidad Nacional de 
             R\'{\i}o Cuarto, R\'{\i}o Cuarto, C\'ordoba, Argentina, \mailsb \and 
Universidad Nacional de C\'ordoba, FaMAF, C\'ordoba, Argentina, \mailsa \and
Consejo Nacional de Investigaciones Cient\'ificas y T\'ecnicas (CONICET), Argentina \and
Saarland University, Saarbr{\"{u}}cken, Germany}

\setcounter{tocdepth}{3}

\maketitle

\begin{abstract}
In this paper we introduce a notion of fault-tolerance distance between 
labeled transition systems. Intuitively, this notion of distance measures the degree of fault-tolerance exhibited by a candidate system.
In practice, there are different kinds of fault-tolerance, here we restrict ourselves to the analysis of
masking fault-tolerance because it is often a highly desirable goal for critical systems. 
Roughly speaking, a system is masking fault-tolerant when it is able to completely mask the faults, 
not allowing these faults to have any observable consequences for the users.  
We capture masking fault-tolerance via a simulation relation, which is accompanied  by a corresponding  game characterization. 
We enrich the resulting games  with quantitative objectives to 
define the notion of masking fault-tolerance distance.
Furthermore, we investigate the basic properties of this notion of masking distance, and 
we prove that it is a directed semimetric.
We have implemented our approach in a prototype tool that  automatically computes the masking distance between a nominal system and a fault-tolerant version of it. 
We have used this tool to measure the masking tolerance of multiple instances of several case studies.


\end{abstract}

\setcounter{page}{1}

\section{Introduction} \label{sec:intro}

Fault-tolerance allows for the construction of systems that are able to 
overcome the occurrence of faults during their execution. 
Examples of fault-tolerant systems can be found everywhere:
communication protocols, hardware circuits, avionic systems, 
cryptographic currencies, etc. 
So, the increasing relevance of critical software in  
everyday life  has led to a renewed interest  in the automatic 
verification of fault-tolerant properties. However, one of the main 
difficulties when reasoning about these kinds of properties is given 
by their quantitative nature, which is true even for non-probabilistic systems. 
A simple example is given by the introduction of redundancy in critical systems. 
This is, by far, one of the most used techniques in fault-tolerance. 
In practice, it is well-known that adding more redundancy to a system increases 
its reliability. Measuring this increment is a central 
issue for evaluating fault-tolerant software, protocols, etc. 
On the other hand, the formal characterization of fault-tolerant properties 
could be an involving task, usually these properties are encoded 
using \emph{ad-hoc} mechanisms as part of a general design.

The usual flow for the design and verification of fault-tolerant systems consists in defining 
a nominal model (i.e., the ``fault-free'' or ``ideal'' program) and afterwards 
extending it with faulty behaviors that deviate from the normal behavior 
prescribed by the nominal model. 
This extended model represents the way in which the system operates 
under the occurrence of faults. 
There are different ways of extending the nominal model, the typical approach is \emph{fault 
injection} \cite{HsuehTI97,IyerNGK10}, that is, the automatic introduction of faults into the model. 
An important property that any extended model has to satisfy is the preservation of the normal behavior under the absence of faults.
In \cite{DemasiCMA17}, we proposed an alternative formal approach 
for dealing with the analysis of fault-tolerance. This approach allows for a fully 
automated analysis and appropriately distinguishes faulty behaviors 
from normal ones. Moreover, this framework is amenable to fault-injection.
In that work, three notions of simulation relations are defined to characterize
\emph{masking}, \emph{nonmasking}, and \emph{failsafe} fault-tolerance, as originally 
defined in \cite{Gartner99}. 

During the last decade, significant progress has been made towards 
defining suitable metrics or distances for diverse types of quantitative 
models including real-time systems \cite{HenzingerMP05}, probabilistic 
models \cite{DesharnaisGJP04}, and metrics for linear and branching 
systems \cite{CernyHR12,AlfaroFS09,Henzinger13,LarsenFT11,ThraneFL10}. 
Some authors have already pointed out that these metrics can be useful to reason 
about the robustness of a system, a notion related to fault-tolerance. 
Particularly,  in \cite{CernyHR12} the traditional notion 
of simulation relation is generalized and three different 
simulation distances between systems are introduced, 
namely \emph{correctness}, \emph{coverage}, and \emph{robustness}.
These are defined using quantitative games with \emph{discounted-sum} and \emph{mean-payoff} objectives. 

In this paper we introduce a notion of fault-tolerance distance between 
labelled transition systems. Intuitively, this distance measures the 
degree of fault-tolerance exhibited by a candidate system. As it was mentioned above, there exist different levels of fault-tolerance, we restrict ourselves to the 
analysis of \emph{masking fault-tolerance} because it is often 
classified as the most benign kind of fault-tolerance and it is a highly 
desirable property for critical systems. 
Roughly speaking, a system is masking fault-tolerant when it is able to completely 
mask the faults, not allowing these faults to have any observable consequences 
for the users. Formally, the system must preserve 
both the safety and liveness properties of the nominal model \cite{Gartner99}.  In contrast to the robustness distance defined in \cite{CernyHR12}, which measures how many unexpected errors are tolerated by the implementation, we consider a specific collection 
of faults given in the implementation and measure how many faults are tolerated by the implementation in such a way that they can be masked by the states. 
We also require that the normal behavior of the specification has to be preserved by the implementation when no faults are present.
In this case, we have a bisimulation between the specification and the 
non-faulty behavior  of the implementation. Otherwise, the distance is $1$.
%
%
That is, $\DeltaMask(N,I)=1$ if and only if the nominal model $N$ and $I\backslash F$ are not bisimilar, where $I\backslash F$ behaves like the implementation $I$ where all actions in $F$ are forbidden ($\backslash$ is Milner's restriction operator).
Thus, we effectively distinguish between the nominal model and its fault-tolerant version and the set of faults taken into account.
	

In order to measure the degree of masking fault-tolerance of a given system, 
we start characterizing masking fault-tolerance via simulation relations between 
two systems as defined in \cite{DemasiCMA17}. The first one acting as a specification 
of the intended behavior (i.e., nominal model) and the 
second one as the fault-tolerant implementation (i.e., the extended model with 
faulty behavior).
The existence of a masking relation implies that the implementation masks the faults.
Afterwards, we introduce a game characterization of 
masking simulation and we enrich the resulting games 
with quantitative objectives to define the notion of 
\emph{masking fault-tolerance distance}, 
where the possible values of the game belong to the interval $[0,1]$. 
The fault-tolerant implementation is masking fault-tolerant
if the value of the game is $0$. Furthermore, the bigger the number, the 
farther the masking distance between the fault-tolerant implementation 
and the specification. Accordingly, a bigger distance remarkably 
decreases fault-tolerance.
Thus, for a given nominal model $N$ and two different fault-tolerant implementations $I_1$ and $I_2$, our distance ensures that $\DeltaMask(N,I_1)<\DeltaMask(N,I_2)$ whenever $I_1$ tolerates more faults than $I_2$.
We also provide a weak version of masking simulation, which makes it possible to
deal with complex systems composed of several interacting components.
We prove that masking distance is a directed semimetric, that is, it satisfies 
two basic properties of any distance, reflexivity and the triangle inequality. 

Finally, we have implemented our approach in a tool that takes as input a nominal model and its fault-tolerant implementation and automatically compute the masking distance between them. 
We have used this tool to measure the masking tolerance of multiple instances of several case 
studies such as a redundant cell memory, a variation of the dining philosophers problem, 
the bounded retransmission protocol, N-Modular-Redundancy, and the Byzantine generals problem. 
These are typical examples of fault-tolerant systems.


The remainder of the paper is structured as follows. 
In Section \ref{sec:background}, we introduce preliminaries notions used 
throughout this paper.
We present in Section \ref{sec:masking_dist} the formal definition 
of masking distance build on quantitative simulation games and we 
also prove its basic properties. 
We describe in Section \ref{sec:experimental_eval} the experimental 
evaluation on some well-known case studies.
In Section \ref{sec:related_work} we discuss the related work. 
Finally, we discuss in Section \ref{sec:conclusions} some conclusions 
and directions for further work.
Full details and proofs can be found in \cite{CastroDDP18}.

\section{Preliminaries} \label{sec:background}

Let us introduce some basic definitions and results on game theory that will be necessary across the paper, the interested reader is referred to \cite{AptG11}.
	
A \emph{transition system} (TS) is a tuple 
$A =\langle S, \Sigma, E, s_0\rangle$, where $S$ is a finite set of states, 
$\Sigma$ is a finite alphabet, $E \subseteq S \times \Sigma \times S$ is a 
set of labelled transitions, and $s_0$ is the initial state. In the following we use $s \xrightarrow{e} s' \in E$ to denote $(s,e,s') \in E$.
Let $|S|$ and $|E|$ denote the number of states and edges, respectively. 
We define $post(s) = \{s' \in S \mid s \xrightarrow{e} s' \in E\}$ as the set of successors of $s$. Similarly, 
$pre(s') = \{s \in S \mid s \xrightarrow{e} s' \in E \}$ as the set of predecessors of $s'$.
Moreover, $post^{*}(s)$ denotes the states which are reachable from $s$. 
Without loss of generality, we require that every state $s$ has a successor, i.e.,
 $\forall s \in S : post(s) \neq \emptyset$. 
A run in a transition system $A$ is an infinite path 
$\rho = \rho_0 \sigma_0 \rho_1 \sigma_1  \rho_2 \sigma_2 \dots \in (S \cdot \Sigma)^{w}$ 
where $\rho_0 = s_0$ and for all $i$, $\rho_i \xrightarrow{\sigma_i} \rho_{i+1} \in E$. From now on, given a tuple
$(x_0,\dots,x_n)$, we denote by $\pr{i}{(x_0,\dots,x_n)}$ its $i$-th projection.

A \emph{game graph} $G$ is a tuple $G = \langle S, S_1, S_2, \Sigma, E, s_0 \rangle$ where $S$, 
$\Sigma$, $E$ and $s_0$ are as in transition systems and $(S_1, S_2)$ is a partition of $S$. 
The choice of the next state is made by Player $1$ (Player $2$) when the current state is in $S_1$ (respectively, $S_2$). A weighted game graph is a game graph along with a weight function $v^G$ 
from $E$ to $\mathbb{Q}$. A run in the game graph $G$ is called a \emph{play}. 
The set of all plays is denoted by $\Omega$.

Given a game graph $G$, a \emph{strategy} for Player $1$ is a function 
$\pi: (S \cdot \Sigma)^{*} S_1 \rightarrow S \times \Sigma$ such that 
for all $\rho_0 \sigma_0 \rho_1 \sigma_1 \dots \rho_i~\in~(S \cdot \Sigma)^{*} S_1$,
we have that if $\pi(\rho_0 \sigma_0 \rho_1 \sigma_1 \dots \rho_i) = (\sigma, \rho)$,
then $\rho_i \xrightarrow{\sigma} \rho \in E$. A strategy for Player $2$ is defined in a similar way. The set of all strategies for Player $p$ is denoted by $\Pi_{p}$.
A strategy for player $p$ is said to be memoryless (or positional) if it can be defined by a mapping $f:S_p \rightarrow E$ such that for all 
$s \in S_p$ we have that
$\pr{0}{f(s)}=s$,
that is, these strategies do not need memory of the past history.
Furthermore, a play $\rho_0 \sigma_0 \rho_1 \sigma_1  \rho_2 \sigma_2 \dots$ conforms to a player $p$ 
strategy $\pi$ if $\forall i \geq 0: (\rho_i \in S_p) \Rightarrow (\sigma_{i}, \rho_{i+1}) = \pi(\rho_0 \sigma_0 \rho_1 \sigma_1 \dots \rho_i)$. The \emph{outcome} of a Player $1$ 
strategy $\pi_{1}$ and a Player $2$ strategy $\pi_2$ is the unique play, named $out(\pi_1, \pi_2)$,  
that conforms to both $\pi_1$ and $\pi_2$.

A \emph{game} is made of a game graph and a boolean or quantitative objective. A \emph{boolean objective} is a function $\Phi: \Omega \rightarrow \{0, 1\}$ and the goal of Player $1$ in a game with objective $\Phi$ is to select a strategy so that the outcome maps to $1$, independently what Player $2$ does. On the contrary, the goal of Player $2$ is 
to ensure that the outcome maps to $0$. Given a boolean objective $\Phi$, a play $\rho$ is \emph{winning} for Player $1$ (resp. Player $2$) 
if $\Phi(\rho) = 1$ (resp. $\Phi(\rho) = 0$). A strategy $\pi$ is a \emph{winning strategy} for Player $p$ if every play conforming to  $\pi$ is winning for Player $p$. We say that a game
with boolean objective is \emph{determined} if some player has a winning strategy, and we say that it is memoryless determined if that winning strategy is memoryless. Reachability games are those games whose objective functions are defined as 
$\Phi(\rho_0 \sigma_0 \rho_1 \sigma_1  \rho_2 \sigma_2 \dots) = (\exists i : \rho_i \in V)$ for some set $V \subseteq S$, a standard result is that reachability games are memoryless determined.

A \emph{quantitative objective} is given by a \emph{payoff} function $f: \Omega \rightarrow \mathbb{R}$ 
and the goal of Player $1$ is to maximize the value $f$ of the play, whereas the goal of 
Player $2$ is to minimize it. For a quantitative objective $f$, the value of the game for a Player $1$ strategy $\pi_1$, 
denoted by $v_1(\pi_1)$, is defined as the infimum over all the values  
resulting from Player $2$ strategies, i.e., $v_1(\pi_1) = \inf_{\pi_2 \in \Pi_2} f(out(\pi_1, \pi_2))$.
The value of the game for Player $1$ is defined as the supremum of the values of all Player $1$ strategies, i.e., $\sup_{\pi_1 \in \Pi_1} v_1(\pi_1)$.
Analogously, the value of the game for a Player $2$ strategy $\pi_2$ and the value of the game 
for Player $2$ are defined as $v_2(\pi_2) = \Sup_{\pi_1 \in \Pi_1} f(out(\pi_1, \pi_2))$ 
and $\inf_{\pi_2 \in \Pi_2} v_2(\pi_2)$, respectively. We say that a game is determined if both values are equal, that is:
$\sup_{\pi_1 \in \Pi_1} v_1(\pi_1) = \inf_{\pi_2 \in \Pi_2} v_2(\pi_2)$. In this case we denote by $val(\mathcal{G})$ the value of game $\mathcal{G}$.
	The following result from \cite{Martin98} characterizes a large set of determined games.
\begin{theorem} Any game with a quantitative function $f$ that is bounded and Borel measurable is determined.
\end{theorem}

\section{Masking Distance} \label{sec:masking_dist}
We start by defining masking simulation. In \cite{DemasiCMA17}, we 
have defined a state-based simulation for masking fault-tolerance, 
here we recast this definition using labelled transition systems. 
	First, let us introduce some concepts needed for defining masking fault-tolerance.
For any vocabulary $\Sigma$, and set of labels $\mathcal{F} = \{F_0, \dots, F_n\}$ not belonging to $\Sigma$, we consider 
$\SigmaF = \Sigma \cup \mathcal{F}$, where $\mathcal{F} \cap \Sigma = \emptyset$. Intuitively, the elements of $\mathcal{F}$ indicate the 
occurrence of a fault in a faulty implementation. Furthermore, sometimes it will be useful to consider the set $\Sigma^i = \{ e^i \mid e \in \Sigma\}$, containing the elements of $\Sigma$ indexed with superscript $i$.
Moreover, for any vocabulary $\Sigma$ we consider $\SigmaM = \Sigma \cup \{M\}$, 
where $M \notin \Sigma$, intuitively, this label is used to identify masking  transitions.

Given a transition system $A = \langle S, \Sigma, E, s_0 \rangle$ over a vocabulary 
$\Sigma$, we denote $A^M = \langle S, \SigmaM, E^M, s_0 \rangle$ 
where $E^M = E \cup \{s \xrightarrow{M} s \mid s \in S\}$.

\subsection{Strong Masking Simulation}

\begin{definition} \label{def:masking_rel}
  Let $A =\langle S, \Sigma, E, s_0\rangle$ and $A' =\langle S', \SigmaF, E', s_0' \rangle$ be  two transition systems.
  $A'$ is \emph{strong masking fault-tolerant} with respect to $A$ if there exists a relation 
$\M \subseteq S \times S'$ (considering $A^M =\langle S, \SigmaM, E, s_0\rangle$ instead of $A$) such that:
\begin{enumerate}[(A)]
	\item $s_0 \mathrel{\M} s'_0$, and
	\item for all $s \in S, s' \in S'$ with $s \mathrel{\M} s'$ and all $e \in \Sigma$ the following holds:

	\begin{enumerate}[(1)]
		\item if $(s \xrightarrow{e} t) \in E$ then
		$\exists\; t' \in S': (s' \xrightarrow{e} t'  \wedge t \mathrel{\M}t')$;

     	\item if $(s' \xrightarrow{e} t') \in E'$ then
     	$\exists \; t \in S: (s \xrightarrow{e} t \wedge t \; \M\; t')$;

     	\item if $(s' \xrightarrow{F} t')$ for some $F \in \mathcal{F}$ then
     	$\exists\; t \in S: (s \xrightarrow{M} t \wedge t \mathrel{\M} t').$
	\end{enumerate}
\end{enumerate}
If such relation exist we denote it by $A \Masking A'$ and say that $A'$ is a \emph{strong masking fault-tolerant implementation} of $A$. 
\end{definition}
We say that state $s'$ is masking fault-tolerant for $s$ when $s~\M~s'$. Intuitively, the definition states that, starting in $s'$, faults can be masked in such a way that the behavior exhibited is the same as that observed when starting from $s$ and executing transitions without faults. 
 In other words, a masking relation ensures that every faulty behavior in the implementation can be simulated by the specification. Let us explain in more detail the above definition.
First, note that conditions $A$, $B.1$, and $B.2$ imply that we have a bisimulation when $A$ and $A'$ do not exhibit faulty behavior.
Particularly, condition $B.1$ says that the normal execution of $A$ can be simulated by an execution of $A'$.  On the other hand, condition $B.2$ says that the implementation does not add normal (non-faulty) behavior. Finally, condition $B.3$ states that every outgoing faulty transition ($F$) from $s'$ must be matched to an outgoing masking transition ($M$) from $s$.


\subsection{Weak Masking Simulation}

For analysing nontrivial systems a weak version of masking simulation relation is needed, 
the main idea is that weak masking simulation abstracts away from
internal behaviour, which is modeled by a special 
action $\tau$.  Note that internal transitions are common in fault-tolerance: the actions performed as part of a fault-tolerant procedure in 
a component are usually not observable by the rest of the system.

The \textit{weak transition relations} ${\Rightarrow} \subseteq S
\times (\Sigma \cup \{\tau\} \cup \{M\} \cup \mathcal{F}) \times S$,
also denoted as $E_W$, considers the \emph{silent} step $\tau$ and is
defined as follow: \\
%
\[\xRightarrow{e} = 
 			 \begin{cases}
			    	(\xrightarrow{\tau})^{*}\circ\xrightarrow{e}\circ(\xrightarrow{\tau})^{*} & 
			    	\text{if } e \in \Sigma,  \\ 
    				(\xrightarrow{e})^{*} & \text{if } e = \tau,  \\
    				\xrightarrow{e} & \text{if } e \in \{M\} \cup \mathcal{F}.\\
 			 \end{cases}\]
The symbol $\circ$ stands for composition of binary relations and 
$(\xrightarrow{\tau})^{*}$ is the reflexive and transitive closure of the 
binary relation $\xrightarrow{\tau}$. 

Intuitively, if $e \notin \{\tau,M\}\cup\mathcal{F}$, then
$s\xRightarrow{e}s'$ means that there is a sequence of zero or more $\tau$
transitions starting in $s$, followed by one transition labelled by
$e$, followed again by zero or more $\tau$ transitions eventually
reaching $s'$.
$s \xRightarrow{\tau} s'$ states that $s$ can transition to $s'$ via
zero or more $\tau$ transitions.
In particular, $s \xRightarrow{\tau} s$ for avery $s$.
For the case in which $e\in\{M\}\cup\mathcal{F}$,
$s\xRightarrow{e}s'$is equivalente to $s\xrightarrow{e}s'$ and hence
no $\tau$ step is allowed before or after the $e$ transition.

\begin{definition} \label{def:weak_mask}
  Let $A =\langle S, \Sigma, E, s_0\rangle$ and $A' =\langle S',
  \SigmaF, E', s_0' \rangle$ be two transition systems with $\Sigma$
  possibly containing $\tau$.  $A'$ is \emph{weak masking fault-tolerant}
  with respect to $A$ if there is a relation $\M \subseteq S
  \times S'$ (considering $A^M$ instead of $A$) such that:
\begin{enumerate}[(A)]
	\item $s_0 \mathrel{\M} s'_0$
	\item for all $s \in S, s' \in S'$ with $s \mathrel{\M} s'$ and all $e \in \Sigma \cup \{\tau\}$ the following holds:

	\begin{enumerate}[(1)]
		\item if $(s \xrightarrow{e} t) \in E$ then 
		$\exists\; t' \in S': (s' \xRightarrow{\text{e}} t' \in E_W' 
		\wedge t\mathrel{\M}t')$;

     	\item if $(s' \xrightarrow{e} t') \in E'$ then  
     	$\exists \; t \in S: (s \xRightarrow{e} t \in E_W 
     	\wedge t \mathrel{\M} t')$;

     	\item if $(s' \xrightarrow{F} t') \in E'$ for some $F \in \mathcal{F}$ then 
     	$\exists\; t \in S: (s \xrightarrow{M} t \in E 
     	\wedge t \mathrel{\M} t').$
	\end{enumerate}
\end{enumerate}
If such relation exists, we denote it by $A \WeakMasking A'$ and say
that $A'$ is a \emph{weak masking fault-tolerant implementation} of
$A$.
\end{definition}

The following theorem makes a strong connection between strong and
weak masking simulation. It states that weak masking simulation
becomes strong masking simulation whenever transition $\xrightarrow{}$
is replaced by $\xRightarrow{}$ in the original automata.

\begin{theorem} \label{thm:weak_thm}
  Let
$A =\langle S, \Sigma, E, s_0\rangle$ and $A' =\langle S', \SigmaF, E', s_0' \rangle$. 
$\M \subseteq S \times S'$ (considering $A^M$ instead of $A$) is a weak masking simulation if and only if:
\begin{enumerate}[(A)]
	\item $s_0 \mathrel{\M} s'_0$, and
	\item for all $s \in S, s' \in S'$ with $s \mathrel{\M} s'$ and all $e \in \Sigma \cup \{\tau\}$ the following holds:

	\begin{enumerate}[(1)]
		\item if $(s \xRightarrow{e} t) \in E_W$ then $\exists\; t' \in S': (s' \xRightarrow{\text{e}} t' \in E_W' 
		\wedge t\mathrel{\M}t')$;

     	\item if $(s' \xRightarrow{e} t') \in E_W'$ then 
     	$\exists \; t \in S: (s \xRightarrow{e} t \in E_W 
     	\wedge t \mathrel{\M} t')$;

     	\item if $(s' \xRightarrow{F} t') \in E_W'$ for some $F \in \mathcal{F}$ then 
     	$\exists\; t \in S: (s \xRightarrow{M} t \in E_W\wedge t \mathrel{\M} t')$
	\end{enumerate}
\end{enumerate}
\end{theorem}

The proof of this theorem is straightforward following the same ideas of 
Milner in \cite{Milner89}.

A natural way to check weak bisimilarity is to \emph{saturate}
the transition system  \cite{FernandezM91,Milner89} and then check strong bisimilarity on the 
saturated transition system. 
Similarly, Theorem~\ref{thm:weak_thm} allows us to compute weak masking simulation by reducing it to 
strong masking simulation. Note that $\xRightarrow{e}$ can be alternatively defined by:
%
%
\[
\dfrac{p \xrightarrow{e} q}{p\xRightarrow{e} q} \hspace{2cm} 
\dfrac{}{p\xRightarrow{\tau} p} \hspace{2cm} 
\dfrac{p\xRightarrow{\tau} p_1 \xRightarrow{e} q_1 \xRightarrow{\tau} q}{p\xRightarrow{e} q}~(e \notin \{M\} \cup \mathcal{F})
\]



As a running example, we consider a memory cell that stores a bit of information and supports reading and writing operations, presented in a state-based form in \cite{DemasiCMA17}. A state in this system maintains the current value of the memory cell ($m=i$, for $i=0,1$), writing allows one to change this value, and reading returns the stored value.  
Obviously, in this system the result of a reading depends on the value stored in the cell. 
Thus, a property that one might associate with this model is that the value read from the cell coincides with that of the last writing performed in the system. 
    
A potential fault in this scenario occurs when a cell unexpectedly loses its charge, and its stored value turns into another one (e.g., it changes from $1$ to $0$ due to charge loss). A typical technique to deal with this situation is \emph{redundancy}: use three memory bits instead of one. Writing operations are performed simultaneously on the three bits. Reading, on the other hand, returns the value that is repeated at least twice in the memory bits; this is known as \emph{voting}. 

\begin{figure}[h] 
\begin{center}
    \includegraphics[scale=0.45]{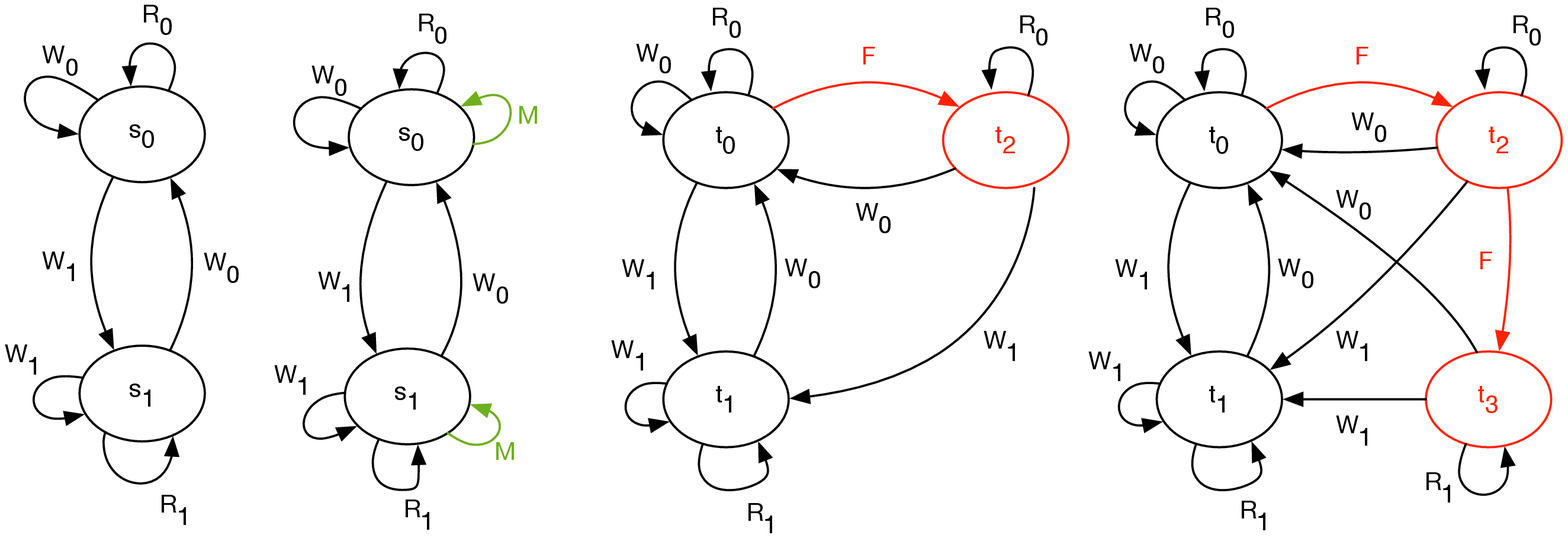} 
    \vspace{-0.8cm}
    \caption{Transition systems for the memory cell.}
    \vspace{-0.5cm}
    \label{figure:exam_1_mem_cell}
\end{center}
\end{figure}

We take the following approach to model this system. Labels $\text{W}_0, \text{W}_1, \text{R}_0,$ and $\text{R}_1$
represent writing and reading operations. Specifically, $\text{W}_0$ (resp. $\text{W}_1$): writes a zero (resp. one) in the memory. $\text{R}_0$ (resp. $\text{R}_1$): reads a zero (resp. one) from the memory.
Figure~\ref{figure:exam_1_mem_cell} depicts four transition systems. 
The leftmost one represents the nominal system for this example (denoted as $A$). 
The second one from the left characterizes the nominal transition system augmented with masking 
transitions, i.e., $A^M$.
The third and fourth transition systems are  fault-tolerant implementations of $A$, named $A'$ and $A''$, respectively. Note that $A'$ contains one fault, while $A''$ considers two faults.   
Both implementations use triple redundancy; intuitively,
state $\text{t}_0$ contains the three bits with value zero and $\text{t}_1$ contains the three bits with value one. 
Moreover, state $\text{t}_2$ is reached when one of the bits was flipped (either $001$, $010$ or $100$).
In  $A''$, state $\text{t}_3$ is reached after a second bit is flipped (either $011$ or $101$ or $110$) starting from state $\text{t}_0$.
It is straightforward to see that there exists a relation of masking fault-tolerance between $A^M$ and $A'$, as it is witnessed by the relation $\M = \{(\text{s}_0, \text{t}_0), (\text{s}_1, \text{t}_1), (\text{s}_0, \text{t}_2)\}$. It is a routine to check that $\M$ satisfies the conditions of Definition \ref{def:masking_rel}.
On the other hand, there does not exist a masking relation between $A^M$ and $A''$ because state $\text{t}_3$ needs to be related to state $\text{s}_0$ in any masking relation. This  state can only be reached by executing faults, which are necessarily masked with $M$-transitions. However, note that, in state  $\text{t}_3$, we can read a $1$ (transition $\text{t}_3 \xrightarrow{\text{R}_1} \text{t}_3$) whereas, in state $\text{s}_0$, we can only read a $0$.
 
\subsection{Masking Simulation Game} \label{subsec:mask_sim_game}

We define a masking simulation game for two transition systems (the
specification of the nominal system and its fault-tolerant
implementation) that captures masking fault-tolerance.  We first
define the masking game graph where we have two players named by
convenience the \emph{refuter} ($\Refuter$) and the \emph{verifier}
($\Verifier$).

\begin{definition} \label{def:strong_masking_game_graph}
  Let $A=\langle S, \Sigma, E, s_0\rangle$ and $A'=\langle S',
  \Sigma_{\mathcal{F}}, E_W', s'_0 \rangle$.
  The \emph{strong masking game graph} $\mathcal{G}_{A^M,A'}
  = \langle S^G, S_R, S_V, \Sigma^G, E^G, {s_0}^G \rangle$ for two
  players is defined as follows:
\begin{itemize}
    \item $\Sigma^G = \SigmaM \cup \SigmaF$
	\item $S^G = (S \times ( \SigmaM^1 \cup \SigmaF^2 \cup\{\#\}) \times S' \times \{ R, V \}) 
	\cup \{\ErrorSt\}$
	\item The initial state is $s_0^G = \langle s_0, \#, s'_0, R \rangle$, where the refuter starts 
	playing
	\item The refuter's states are $S_R = \{ (s, \#, s', R) \mid s \in S \wedge s' \in S' \} 
	\cup \{\ErrorSt\}$
	\item The verifier's states are $S_V = \{ (s, \sigma, s', V) \mid s \in S \wedge s' \in S' \wedge \sigma \in \Sigma^G\setminus\{M\}\}$
\end{itemize}
and $E^G$ is the minimal set satisfying:
\begin{itemize}
	\item $\{ (s, \#, s', R) \xrightarrow{\sigma} (t, \sigma^{1}, s', V) \mid \exists\;\sigma \in \Sigma: s \xrightarrow{\sigma} t \in E \} \subseteq E^G$,

	\item $\{ (s, \#, s', R) \xrightarrow{\sigma} (s, \sigma^{2}, t', V)  \mid \exists\;\sigma \in \SigmaF: s' \xrightarrow{\sigma} t' \in E' \} \subseteq E^G$,

	\item $\{ (s, \sigma^2, s', V) \xrightarrow{\sigma} (t, \#, s', R) \mid \exists\;\sigma \in \Sigma: s \xrightarrow{\sigma} t \in E \} \subseteq E^G$,

	\item $\{ (s, \sigma^1, s', V) \xrightarrow{\sigma} (s, \#, t', R) \mid \exists\;\sigma \in \Sigma: s' \xrightarrow{\sigma} t' \in E' \} \subseteq E^G$,

	\item $\{ (s, F_i^2, s', V) \xrightarrow{M} (t, \#, s', R) \mid \exists\;s \xrightarrow{M} t \in E^M \} \subseteq E^G$, for any $F_i \in \mathcal{F}$ 

    \item If there is no outgoing transition from some state $s$ then transitions $s \xrightarrow{\sigma} \ErrorSt$ and $\ErrorSt \xrightarrow{\sigma} \ErrorSt$ for every $\sigma \in \Sigma$, are added.
\end{itemize}
\end{definition}

The intuition of this game is as follows. 
The refuter chooses transitions of either the specification or the implementation to play,
and the verifier tries to match her choice, this is similar to the bisimulation game \cite{Stirling99}. However, when the refuter chooses a fault, the verifier must match it with a masking transition ($M$). The intuitive reading of this is that the fault-tolerant implementation masked the fault in such a way that the occurrence of this fault cannot be noticed from the users' side. $\Refuter$ wins if the game reaches the error state, i.e., $\ErrorSt$. On the other hand, $\Verifier$ wins when $\ErrorSt$ is not reached during the game. (This is basically a reachability game \cite{Jurd11}). We 
say $\text{Ver}(v)$ (resp. $\text{Ref}(v)$) if $v$ is a verifier's node  (resp. refuter's node).

A \emph{weak masking game graph} $\mathcal{G}^W_{A^M,A'}$ is defined
in the same way as the strong masking game graph in
Def.~\ref{def:strong_masking_game_graph}, with the exception that
$\SigmaM$ and $\SigmaF$ may contain $\tau$, and the set of labelled
transitions (denoted as $E_W^G$) is now defined using the weak
transition relations (i.e., $E_W$ and $E_W'$) from the respective
transition systems.

\begin{figure} [h]
\begin{center}
    \vspace{-0.6cm}
    \includegraphics[scale=0.49]{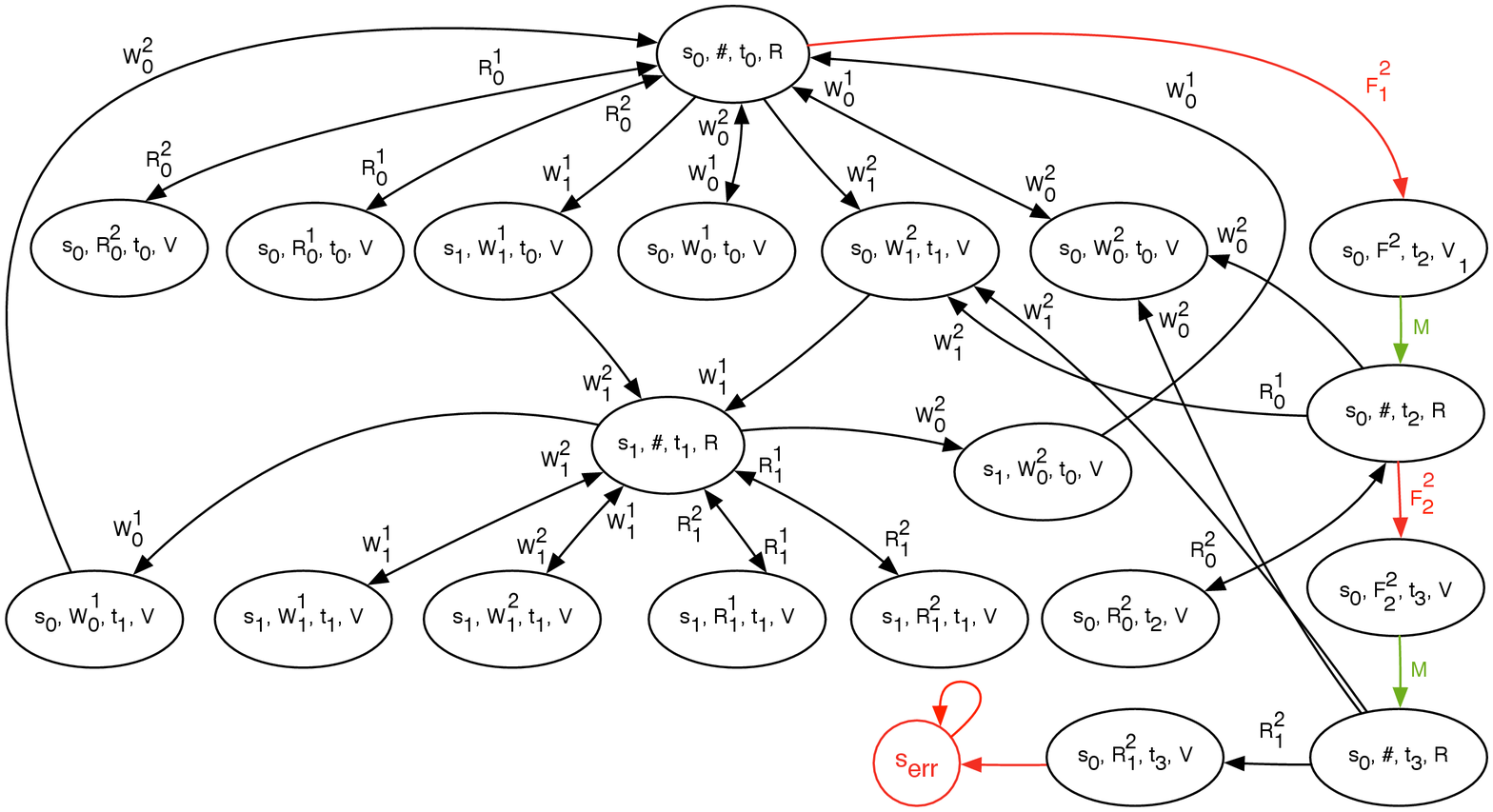} 
    \vspace{-0.7cm}
    \caption{Part of the masking game graph for memory cell model with two faults}
    \label{figure:exam_2_mem_cell_gg_two_faults}
     \vspace{-0.8cm}
\end{center}
\end{figure}

Figure~\ref{figure:exam_2_mem_cell_gg_two_faults} shows a part 
of the strong masking game graph for the running example considering the transition 
systems $A^M$ and $A''$. 
We can clearly observe on the game graph that the verifier cannot mimic the 
transition $(s_0, \#, t_3, R) \xrightarrow{R_1^2} (s_0, R_1^2, t_3, V)$
selected by the refuter which reads a $1$ at state $t_3$ on the fault-tolerant 
implementation. This is because the verifier can only read a $0$ at state $s_0$. 
Then, the $\ErrorSt$ is reached and the refuter wins.

As expected, there is a strong masking simulation between $A$ and $A'$
if and only if the verifier has a winning strategy in
$\mathcal{G}_{A^M,A'}$.

\begin{theorem} \label{thm:wingame_strat}
  Let $A=\langle S, \Sigma, E, s_0\rangle$ and $A'=\langle S', \SigmaF, E', s_0' \rangle$.
  $A \Masking A'$ iff the verifier has a winning strategy for the strong masking game graph $\mathcal{G}_{A^M,A'}$.
\end{theorem}

By Theorems~\ref{thm:weak_thm} and~\ref{thm:wingame_strat}, the result replicates for weak masking game.

\begin{theorem} \label{thm:weak_wingame_strat}
  Let $A=\langle S, \Sigma \cup \{\tau\}, E, s_0\rangle$ and
  $A'=\langle S', \SigmaF \cup \{\tau\}, E', s_0' \rangle$.
  $A \WeakMasking A'$ iff the verifier has a winning strategy for the
  weak masking game graph $\mathcal{G}^W_{A^M,A'}$.
\end{theorem}

Using the standard properties of reachability games we get the following property.

\begin{theorem}
  For any $A$ and $A'$, the strong (resp.\ weak) masking game graph
  $\mathcal{G}_{A^M, A'}$ (resp.\ $\mathcal{G}^W_{A^M, A'}$) is
  determined.
  Furthermore, the strong (resp.\ weak) masking game graph can be
  determined in time $O(|E^G|)$ (resp.\ $O(|E_W^G|)$).
\end{theorem}

The set of winning states for the refuter can be defined in a standard way from the error state \cite{Jurd11}. We adapt ideas in \cite{Jurd11} to our setting. For $i,j\geq 0$,  sets $U^j_i$ are defined as follows:

\begin{align}
  U^0_i =& U^j_0 = \emptyset \label{def:of:Uji}\\
  U_1^1 =&  \{\ErrorSt\},
  \hspace{10cm} {}\notag
\end{align}
\begin{align*}
  U_{i+1}^{j+1} &=
    \{v' \mid Ref(v') \wedge post(v') \cap U_{i+1}^j \neq \emptyset\} \\
    &\textstyle \hspace{-1em}{}\cup
    \{v' \mid Ver(v') \wedge post(v') \subseteq \bigcup_{j'\leq j} U_{i+1}^{j'} \wedge post(v') \cap U^j_{i+1} \neq \emptyset \wedge \pi_2(v') \notin \mathcal{F} \} \\
    &\textstyle \hspace{-1em}{}\cup
    \{v' \mid  Ver(v') \wedge post(v') \subseteq \bigcup_{i'\leq i, j' \leq j}U_{i'}^{j'} \wedge post(v') \cap U^j_{i} \neq \emptyset \wedge \pi_2(v') \in \mathcal{F} \}
\end{align*}
%
%
then $U^k = \bigcup_{i \geq 0} U^k_i$ and $U = \bigcup_{k \geq 0} U^k$. Intuitively, the subindex $i$ in $U^k_i$ indicates that $\ErrorSt$ is reach after at most $i-1$ faults occurred.
The following lemma is straightforwardly proven using standard techniques of reachability games \cite{AlfaroHK07}.

\begin{lemma} \label{lemma:RefWinStrat} The refuter has a winning strategy in $\mathcal{G}_{A^M, A'}$ (or $\mathcal{G}^W_{A^M, A'}$) iff $s_{init} \in U^k$, for some $k$.
\end{lemma}

\subsection{Quantitative Masking}
In this section, we extend the strong masking simulation game introduced above 
with quantitative objectives to define the notion of masking fault-tolerance distance.
Note that we use the attribute ``quantitative'' in a non-probabilistic sense.
\begin{definition}	
  For transition systems $A$ and $A'$, the \emph{quantitative strong masking game graph} $\mathcal{Q}_{A^M, A'} = \langle S^G, S_R, S_V, \Sigma^G, E^G, s_{0}^G, v^G \rangle$ is defined as follows:
\begin{itemize}
\item
  $\mathcal{G}_{A^M, A'}=\langle S^G, S_R, S_V, \Sigma^G,  E^G, s_{0}^G \rangle$ is defined as in Definition~\ref{def:strong_masking_game_graph},
\item
  $ v^G(s \xrightarrow{e} s') = (\chi_{\mathcal{F}} (e), \chi_{s_{err}}(s'))$

\end{itemize}
where $\chi_{\mathcal{F}}$ is the characteristic function over 
set $\mathcal{F}$, returning $1$ if $e \in \mathcal{F}$ and $0$ otherwise, and $\chi_{s_{err}}$ is the characteristic function over the singleton set $\{s_{err}\}$.
\end{definition}

Note that the cost function returns a pair of numbers instead of a
single number. It is direct to codify this pair into a number, but we
do not do it here for the sake of clarity.  We remark that the
\emph{quantitative weak masking game graph} $\mathcal{Q}^W_{A^M, A'}$
is defined in the same way as the game graph defined above but using
the weak masking game graph $\mathcal{G}^W_{A^M, A'}$ instead of
$\mathcal{G}_{A^M, A'}$




Given a quantitative strong masking game graph with the weight function $v^G$ and a play 
$\rho = \rho_0 \sigma_0 \rho_1 \sigma_1 \rho_2, \ldots$, for all $i \geq 0$, let 
$v_i = v^G(\rho_i \xrightarrow{\sigma_i} \rho_{i+1})$.
We define the \emph{masking payoff function} as follow: 
\[
\FMask(\rho) = \lim_{n \rightarrow \infty}  \frac{\pr{1}{v_n}}{1+ \sum^{n}_{i=0} \pr{0}{v_i}},
\]
which is proportional to the inverse of the number of masking movements made by the verifier. To see this, note that the numerator of $\frac{\pr{1}{v_n}}{1+ \sum^{n}_{i=0} \pr{0}{v_i}}$ will be $1$
when we reach the error state, that is, in those paths not reaching the error state this formula returns $0$. Furthermore, if the error state is reached, then the denominator will count the number of fault transitions
taken until the error state. All of them, except the last one, were masked successfully.  The last fault, instead, while attempted to be masked by the verifier, eventually leads to the error state.
That is, the transitions with value $1$ are those corresponding to faults.  The others are mapped to $0$.
Notice also that if $\ErrorSt$ is reached in $v_n$ without the occurrence of any fault, the nominal part of the implementation does not match the nominal specification, in which case $\frac{\pr{1}{v_n}}{1+ \sum^{n}_{i=0} \pr{0}{v_i}}=1$.
Then, 
the refuter wants to maximize the value of any run, that is, she will try to execute  faults leading to the state $\ErrorSt$. 
In contrast, the verifier wants to avoid $\ErrorSt$ and then she will try to mask faults with actions that take her away from the error state. 
More precisely, the value of the quantitative strong masking game for the refuter is defined as  $val_R(\mathcal{Q}_{A^M,A'}) = \Sup_{\pi_R \in \Pi_R} \; \Inf_{\pi_V \in \Pi_V} \FMask(out(\pi_R, \pi_V))$. Analogously, the value of the game for the verifier is defined as $val_V(\mathcal{Q}_{A^M,A'}) = \Inf_{\pi_V \in \Pi_V} \; \Sup_{\pi_R \in \Pi_R} \FMask(out(\pi_R, \pi_V))$. Then, we define the value of the quantitative strong masking game, denoted by $\val(\mathcal{Q}_{A^M,A'})$, as the value 
of the game either for the refuter or the verifier, i.e., $\val(\mathcal{Q}_{A^M,A'}) = val_R(\mathcal{Q}_{A^M,A'}) = val_V(\mathcal{Q}_{A^M,A'})$. This can be done because 
quantitative strong masking games are determined as we prove below in Theorem~\ref{thm:mask_game_det}. 

\begin{definition} \label{def:mask_dist}
  Let  $A$ and $A'$ be transition systems. The \emph{strong masking distance} between $A$ and $A'$, denoted by $\DeltaMask(A, A')$ is defined as:
$\DeltaMask(A, A') = \val(\mathcal{Q}_{A^M,A'}).$
\end{definition}

We would like to remark that the \emph{weak masking distance} $\DeltaMask^W$ is defined in the same way 
for the quantitative weak masking game graph $\mathcal{Q}^W_{A^M,A'}$.  Roughly speaking, we are interesting on 
measuring the number of faults that can be masked. The value of the game 
is essentially determined by the faulty and masking labels on the game graph and how 
the players can find a strategy that leads (or avoids) the state $\ErrorSt$, independently if 
there are or not silent actions.
	
In the following, we state some basic properties of this kind of games. 
As already anticipated, quantitative strong masking games are determined:


\begin{theorem} \label{thm:mask_game_det}
  For any quantitative strong masking game $\mathcal{Q}_{A^M, A'}$ with payoff function $\FMask$,
  \[\textstyle
  \Inf_{\pi_V \in \Pi_V} \; \Sup_{\pi_R \in \Pi_R} \FMask(out(\pi_R, \pi_V)) = \Sup_{\pi_R \in \Pi_R} \;  \Inf_{\pi_V \in \Pi_V} \FMask(out(\pi_R, \pi_V))\]
\end{theorem}


The value of the quantitative strong masking game can be calculated as stated below.

\begin{theorem} \label{thm:quant_game}
  Let $\mathcal{Q}_{A^M,A'}$ be a quantitative strong masking
  game.
  Then, $\val(\mathcal{Q}_{A^M,A'}) = \frac{1}{w}$, with
  $w = \min \{ i \mid \exists j : s_{init} \in U^j_i \}$, whenever
  $s_{init} \in U$, and $\val(\mathcal{Q}_{A^M,A'})=0$ otherwise, where
  sets $U^j_i$ and $U$ are defined in equation~(\ref{def:of:Uji}).
\end{theorem}
Note that the sets $U^j_i$ can be calculated using a bottom-up breadth-first search from the error state. Thus, the strategies for the refuter and the verifier can be defined using these 
sets, without taking into account the history of the play. That is, we have the following theorems:
\begin{theorem} \label{thm:memoryless}
  Players $\Refuter$ and $\Verifier$ have memoryless winning strategies for $\mathcal{Q}_{A^M,A'}$.
\end{theorem}
Theorems~\ref{thm:mask_game_det}, \ref{thm:quant_game},
and~\ref{thm:memoryless} apply as well to $\mathcal{Q}^W_{A^M,A'}$.
The following theorem states the complexity of determining the value
of the two types of games.

\begin{theorem}
  The quantitative strong (weak) masking game can be determined in 
time $O(|S^G| + |E^G|)$ (resp. $O(|S^G| + |E_{W}^{G}|)$).
\end{theorem}


By using $\mathcal{Q}^W_{A^M,A'}$ instead of $\mathcal{Q}_{A^M,A'}$ in
Definition~\ref{def:mask_dist}, we can define the \emph{weak masking
  distance} $\DeltaMask^W$.  The next theorem states that, if $A$ and
$A'$ are at distance $0$, there is a strong (or weak) masking
simulation between them.

\begin{theorem}\label{theorem:ref}
  For any transition systems $A$ and  $A'$, then
  \begin{inparaenum}[(i)]
  \item  $\DeltaMask(A,A') = 0$ iff $A \Masking A'$, and
   \item $\DeltaMask^W(A,A') = 0$ iff $A \WeakMasking A' $.
  \end{inparaenum}
\end{theorem}
This follows from Theorem \ref{thm:quant_game}.
Noting that $A \Masking A$ (and $A \WeakMasking A$) for 
any transition system $A$, we obtain that $\DeltaMask(A,A)=0$ (resp.\ $\DeltaMask^W(A,A)=0$) by Theorem \ref{theorem:ref}, i.e., 
both distance are reflexive.

For our running example, the masking distance is $1/3$ 
with a redundancy of $3$ bits and considering two faults. 
This means that only one fault can be masked by this implementation.
We  can  prove a version of the triangle inequality for our notion of distance. 

\begin{theorem}	\label{thm:triang_ineq}
  Let $A = \langle S, \Sigma, E, s_0 \rangle$, $A' = \langle S', \Sigma_{\mathcal{F'}}, E', s'_0 \rangle$,  and  $A'' = \langle S'', \Sigma_{\mathcal{F''}},\\E'', s''_0 \rangle$ be transition systems such that $\mathcal{F}' \subseteq \mathcal{F}''$.
  Then $\DeltaMask(A,A'') \leq \DeltaMask(A,A') + \DeltaMask(A', A'')$ and
  $\DeltaMask^W(A,A'') \leq \DeltaMask^W(A,A') + \DeltaMask^W(A', A'').$
\end{theorem}

Reflexivity and the triangle inequality imply that both masking distances are directed semi-metrics \cite{CharikarMM06,AlfaroMRS08}.  Moreover, it is interesting to note that the triangle inequality property has practical applications. When developing critical software is quite common to develop a first version of the software taking into account some possible anticipated faults. 
Later, after testing and running of the system, more plausible faults could be observed. 
Consequently, the system is modified with additional fault-tolerant capabilities to be able 
to overcome them. 
Theorem \ref{thm:triang_ineq} states that incrementally measuring the masking distance between these different versions of the software provides an upper bound to the actual distance between the nominal system and its last fault-tolerant version. That is, if the sum of the distances obtained between the different versions is a small number, then we can ensure that the final system will 
exhibit an acceptable masking tolerance to faults w.r.t. the nominal system.


\section{Experimental Evaluation} \label{sec:experimental_eval}

The approach described in this paper have been implemented in a tool in \textsf{Java} called \MaskD: 
Masking Distance Tool \cite{MaskD}. 
\MaskD~takes as input a nominal model and its fault-tolerant implementation, 
and produces as output the masking distance between them. 
The input models are specified using the guarded command language introduced in \cite{AroraGouda93}, a simple programming language common for describing fault-tolerant algorithms. 
More precisely, a program is a collection of processes, where each process is composed of a collection of actions of the style: $Guard \rightarrow Command$, where $Guard$ is a boolean condition over the actual state of the program and $Command$ is a collection of basic assignments. These syntactical constructions are called actions. The language also allows user to label 
an action as internal (i.e., $\tau$ actions). Moreover, usually some actions are used to represent faults. 
The tool has several additional features, for instance it can print the traces to the error state or start a simulation from the initial state.

We report on Table~\ref{table:results} the results of the masking distance for multiple instances 
of several case studies. These are: a Redundant Cell Memory (our running example), N-Modular Redundancy (a standard example of fault-tolerant system \cite{ShoomanBook}), a variation of the Dining Philosophers problem \cite{Dijkstra71}, the Byzantine 
Generals problem introduced by Lamport et al. \cite{LamportSP82}, and the Bounded 
Retransmission Protocol (a well-known example of fault-tolerant  protocol \cite{GrooteP96}).

Some words are useful to interpret the results. For the case of a $3$ bit memory the masking distance is $0.333$, the main reason for this is that the faulty model in the worst case is only able to mask $2$ faults (in this example, a fault is an unexpected change of  a bit value) before failing to replicate the nominal behaviour (i.e. reading the majority value), thus the result comes from the definition of masking distance and taking into account the occurrence of two faults. The situation is similar for the other instances of this problem with more redundancy.

N-Modular-Redundancy consists of N systems, in which these perform a process and that results are processed by a majority-voting system to produce a single output. 
Assuming a single perfect voter, we have evaluated this case study for different numbers of modules.
Note that the distance measures for this case study are similar to the memory example. 

%
%
For the dining philosophers problem we have adopted the odd/even philosophers implementation (it prevents from deadlock), i.e.,  there are 
$n-1$ \emph{even} philosophers that pick the right fork first, and $1$ \emph{odd} philosopher that picks the left fork first. The fault we consider in this case occurs when
an even philosopher behaves as an odd one, this could be the case of a byzantine fault. For two philosophers the masking distance is $0{.}5$ since a single fault leads to a deadlock, when more philosophers are added this distance becomes smaller. 

Another interesting example of a fault-tolerant system is the Byzantine generals problem, introduced originally by Lamport et al. \cite{LamportSP82}. This is a consensus problem, where we have a general with $n-1$ lieutenants. The communication between the general and his lieutenants is performed through messengers. The general may decide to attack an enemy city or to retreat; then, he sends the order to his lieutenants. Some of the lieutenants might be traitors. 
We assume that the messages are delivered correctly and all the lieutenants can communicate directly with each other. In this scenario they can recognize who is sending a message. Faults can convert loyal lieutenants into traitors (byzantines faults). As a consequence, traitors might deliver false messages or perhaps they avoid sending a message that they received. The loyal lieutenants must agree on attacking or retreating after $m + 1$ rounds of communication, where $m$ is the maximum numbers of traitors.  


The Bounded Retransmission Protocol (BRP) is a well-known industrial
case study in software verification.
While all the other case studies were treated as toy examples and analyzed with $\DeltaMask$, the BRP was modeled closer to the implementation following~\cite{GrooteP96}, considering the different components (sender, receiver, and models of the channels).  To analyze such a complex model we have used instead the weak masking distance $\DeltaMask^W$.
We have calculated the masking distance for the bounded retransmission protocol with $1$, $3$ and $5$ chunks, denoted BRP(1), BRP(3) and BRP(5), respectively. 
We observe that the distance values are not affected by the number of chunks 
to be sent by the protocol. This is expected because the masking distance depends on 
the redundancy added to mask the faults, which in this case, depends on the number of 
retransmissions.

We have run our experiments on a MacBook Air with Processor 1.3 GHz Intel Core i5 and 
a memory of 4 Gb. The tool and case studies for reproducing the results are available in the tool repository.

\begin{table}[t]
  \centering\noindent%
  \caption{Results of the masking distance for the case studies.}
  \label{table:results}
  \scalebox{0.9}{
    \begin{tabular}{|c|c|c|c|}
      \hline
      \multirow{2}{*}{Case Study} & \multirow{2}{*}{Redundancy} & Masking  & \multirow{2}{*}{Time}  \\
      & & Distance & \\
      \hline
      \multirow{4}{*}{Memory} 	        & $3$ bits & $0.333$ & $0.7s$ \\ \cline{2-4}
      & $5$ bits & $0.25$ & $1.5s$ \\ \cline{2-4}
      & $7$ bits & $0.2$ & $27s$  \\ \cline{2-4}
      & $9$ bits & $0.167$ & $34m33s$ \\ \cline{2-4}
      \hline
      \multirow{4}{*}{\parbox{6em}{\centering N-Modular Redundancy}} 	        & $3$ modules & $0.333$ & $0.3s$ \\ \cline{2-4}
      & $5$ modules & $0.25$ & $0.5s$ \\ \cline{2-4}
      & $7$ modules & $0.2$ & $31.7s$  \\ \cline{2-4}
      & $9$ modules & $0.167$ & $115m$  \\ \cline{2-4}
      \hline
      \multirow{4}{*}{Philosophers}    & $2$ phils & $0.5$ & $0.3s$  \\ \cline{2-4}
      & $3$ phils & $0.333$ & $0.6s$ \\ \cline{2-4}
      & $4$ phils & $0.25$ & $7.1s$  \\ \cline{2-4}
      & $5$ phils & $0.2$ & $13m.53s$ \\ \cline{2-4}
      \hline
      \multirow{2}{*}{Byzantines}    
      & $3$ generals & $0.5$ & $0.5s$  \\ \cline{2-4}
      & $4$ generals & $0.333$ & $2s$ \\ \cline{2-4}
      \hline
    \end{tabular}
    \qquad
    \begin{tabular}{|c|c|c|c|}
      \hline
      \multirow{2}{*}{Case Study} & \multirow{2}{*}{Redundancy} & Masking  & \multirow{2}{*}{Time}  \\
      & & Distance & \\
      \hline
      \multirow{4}{*}{BRP(1)}       	& $1$ retransm. & $0.333$ & $1.2s$ \\ \cline{2-4}
      & $3$ retransm. & $0.2$ & $1.4s$  \\ \cline{2-4}
      & $5$ retransm. & $0.143$ & $1.5s$  \\ \cline{2-4}
      & $7$ retransm. & $0.111$ & $2.1s$  \\ \cline{2-4}
      \hline
      \multirow{4}{*}{BRP(3)}       	& $1$ retransm. & $0.333$ & $5.5s$ \\ \cline{2-4}
      & $3$ retransm. & $0.2$ & $14.9s$  \\ \cline{2-4}
      & $5$ retransm. & $0.143$ & $1m28s$  \\ \cline{2-4}
      & $7$ retransm. & $0.111$ & $4m40s$  \\ \cline{2-4}
      \hline
      \multirow{4}{*}{BRP(5)}       	& $1$ retransm. & $0.333$ & $6.7s$ \\ \cline{2-4}
      & $3$ retransm. & $0.2$ & $32s$  \\ \cline{2-4}
      & $5$ retransm. & $0.143$ & $1m51s$  \\ \cline{2-4}
      & $7$ retransm. & $0.111$ & $6m35s$ \\ \cline{2-4}
      \hline \multicolumn{4}{c}{}\\ \multicolumn{4}{c}{}\\
  \end{tabular}}
\end{table}

\section{Related Work} \label{sec:related_work}
In recent years, there has been a growing interest in the quantitative 
generalizations of the boolean notion of correctness and the 
corresponding quantitative verification questions \cite{BokerCHK14,CernyHR12,Henzinger10,Henzinger13}.
The framework described in \cite{CernyHR12} is the closest related work to our approach. 
The authors generalize the traditional notion of 
simulation relation to three different versions of simulation distance: \emph{correctness}, \emph{coverage}, and \emph{robustness}.
These are defined using quantitative games with \emph{discounted-sum} 
and \emph{mean-payoff} objectives, two well-known cost functions.
Similarly to that work, we also consider distances between purely 
discrete (non-probabilistic, untimed) systems.
Correctness and coverage distances are concerned with the nominal part of the systems, 
and so faults play no role on them. On the other hand, robustness distance measures how many unexpected errors can be performed by the implementation in such a way that the resulting behavior is tolerated by the specification. So, it can be used to analyze the resilience of the implementation. Note that, robustness
distance can only be applied to correct implementations, that is, implementations that preserve the behavior of the specification but perhaps do not cover all its behavior. As noted in~\cite{CernyHR12}, bisimilarity sometimes implies a distance of $1$. In this sense a greater grade of robustness (as defined in~\cite{CernyHR12}) is achieved by pruning critical points from the specification. Furthermore, the  errors considered in that work are transitions mimicking the original ones but with different labels. In contrast to this, 
 in our approach we consider that faults are injected into the fault-tolerant 
 implementation, where their behaviors are not restricted by the nominal system. 
 This follows the idea of model extension in fault-tolerance where faulty behavior is added 
 to the nominal system. Further, note that when no faults are present, the masking distance between the specification and the implementation is $0$ when they are bisimilar, and it is $1$ otherwise.
It is useful to note that robustness distance of~\cite{CernyHR12} is not reflexive. We believe that all these definitions of distance between systems capture different notions useful for software development, and they can be used together, in a complementary way, to obtain an in-depth
evaluation of fault-tolerant implementations.

\section{Conclusions and Future Work}\label{sec:conclusions}
In this paper, we  presented a notion of masking fault-tolerance 
distance between systems built on a characterization of masking 
tolerance via simulation relations and a corresponding game 
representation with quantitative objectives. 
Our framework is well-suited to support engineers for the analysis and 
design of fault-tolerant systems. More precisely, we have defined a 
computable masking distance function such that an engineer 
can measure the masking tolerance of a given 
fault-tolerant implementation, i.e., the number of faults that can be masked. 
Thereby, the engineer can measure and compare the masking fault-tolerance 
distance of alternative fault-tolerant implementations, and select one that 
fits best to her preferences.

There are many directions for future work. 
We have only defined a notion of fault-tolerance distance 
for masking fault-tolerance, similar notions of distance can be defined
for  other levels of fault-tolerance like failsafe and 
non-masking, we leave this as a further work.


\newpage
\bibliographystyle{plain}
\bibliography{references}

\newpage

\appendix

\section{Definitions}

\begin{definition} \label{def:weak_masking_game_graph} Given two transition systems $A=\langle S, \Sigma, E, s_0\rangle$ and $A'=\langle S', \Sigma_{\mathcal{F}}, E_W', s'_0 \rangle$
(where $\Sigma$ and $\SigmaF$ possible contains the distinguished \textit{silent} action $\tau$), 
we define the \emph{weak masking game graph} $\mathcal{G}^W_{A^M,A'} = \langle S^G, S_R, S_V, \Sigma^G,  E_W^G, {s_0}^G \rangle$ 
for two players as follows:
\begin{itemize}
    \item $\Sigma^G = \SigmaM \cup \SigmaF \cup \{\tau\}$
	\item $S^G = (S \times ( \SigmaM^1 \cup \SigmaF^2 \cup \{\tau\} \cup\{\#\}) \times S' \times \{ R, V \}) 
	\cup \{\ErrorSt\}$
	\item The initial state is $s_0^G = \langle s_0, \#, s'_0, R \rangle$, where the refuter starts 
	playing
	\item The refuter's states are $S_R = \{ (s, \#, s', R) \mid s \in S \wedge s' \in S' \} 
	\cup \{\ErrorSt\}$
	\item The verifier's states are $S_V = \{ (s, \sigma, s', V) \mid s \in S \wedge s' \in S' \wedge \sigma \in \Sigma^G\setminus\{M\}\}$
\end{itemize}
and $E_W^G$ is the minimal set satisfying:
\begin{itemize}
	\item $\{ (s, \#, s', R) \xrightarrow{\sigma} (t, \sigma^{1}, s', V) \mid \exists\;\sigma \in \Sigma: s \xRightarrow{\sigma} t \in E_W \} \subseteq E_W^G$,

	\item $\{ (s, \#, s', R) \xrightarrow{\sigma} (s, \sigma^{2}, t', V)  \mid \exists\;\sigma \in \SigmaF: s' \xRightarrow{\sigma} t' \in E_W' \} \subseteq E_W^G$,

	\item $\{ (s, \sigma^2, s', V) \xrightarrow{\sigma} (t, \#, s', R) \mid \exists\;\sigma \in \Sigma: s \xRightarrow{\sigma} t \in E_W \} \subseteq E_W^G$,

	\item $\{ (s, \sigma^1, s', V) \xrightarrow{\sigma} (s, \#, t', R) \mid \exists\;\sigma \in \Sigma: s' \xRightarrow{\sigma} t' \in E_W' \} \subseteq E_W^G$,

	\item $\{ (s, F_i^2, s', V) \xrightarrow{M} (t, \#, s', R) \mid \exists\;s \xRightarrow{M} t \in E^M \} \subseteq E_W^G$, for any $F_i \in \mathcal{F}$ 

    \item If there is no outgoing transition from some state $s$ then transitions $s \xrightarrow{\sigma} \ErrorSt$ and $\ErrorSt \xrightarrow{\sigma} \ErrorSt$ for every $\sigma \in \Sigma$, are added.
\end{itemize}
\end{definition}

\section{Proofs of Properties}

\noindent
\textbf{Proof of Lemma \ref{lemma:RefWinStrat}.} 
The refuter has a winning strategy in $\mathcal{G}_{A^M, A'}$ (or $\mathcal{G}^W_{A^M, A'}$) iff $s_{init} \in U^k$, for some $k$.

\begin{proof} 
	$\Rightarrow$) Suppose that the Refuter has a winning strategy namely $\pi$ and that $s_{init} \notin U^k_i$ for any $i$ and $k$. This means that $\pi(s_{init})$ returns a node $v$ 
such that $v \notin U^k_i$ (for any $i$ and $k$) (by definition of $U^k_i$), and from there
the Verifier can select a node $v' \notin U^k_i$ (for any $i$ and $k$), and 
again this can be repeated forever. Therefore, the play never reaches $s_{err}$, which means that 
the Verifier wins and that leads to a contradiction.	

$\Leftarrow$) Consider $s_{init} \in U^k$ for some $k$, where we have that $s_{init} \in U^k_i$ for some $i$ by definition. Any winning strategy for the refuter is simple, for any $v \in U^j_i$, 
$\pi(v) = v'$ being $v'$ some node in $U^{j-1}_{i}$ 
which exists by definition. Since $s_{init} \in U^k$ and the
Refuter has to play, then the play will reach in $j-1$ steps the set $U^1$, i.e., the $\ErrorSt$ state. \\

\noindent
The proof also apply for the weak masking game graph $\mathcal{G}^W_{A^M, A'}$. 
\end{proof}\\

\noindent
\textbf{Proof of Theorem \ref{thm:wingame_strat}.}
  Let $A=\langle S, \Sigma, E, s_0\rangle$ and $A'=\langle S', \SigmaF, E', s_0' \rangle$.
  $A \Masking A'$ iff the verifier has a winning strategy for the strong masking game graph $\mathcal{G}_{A^M,A'}$.\\

\begin{proof} 
	$\Rightarrow)$ Suppose that $A \Masking A'$, then there exists a masking simulation 
	$\M \subseteq S \times S'$ by Definition \ref{def:masking_rel}.
	Then, the strategy of the verifier is constructed as follows, 
for states $(s, \sigma^i, s', V)$ with $s \; \M \; s'$ and $\sigma \notin \mathcal{F}$, the strategy selects either a transition $s \xrightarrow{\sigma} w$ or $s' \xrightarrow{\sigma} w'$ depending if $i=1$ or $i=2$, respectively. In case of $\sigma \in \mathcal{F}$, then the strategy 
returns the transition $(s, F_j^2, s', V) \xrightarrow{M} (s, \#, s', R)$, for any $F_j \in \mathcal{F}$. 
This can be done since $s \; \M \; s'$. 
Furthermore, in any case we have $w \; \M \; s'$, 
$s\;\M\;w'$ or $s\;\M\; s'$, respectively. 
Thus, the strategy can be applied again for any movement of the Refuter. 
Summing up, the Verifier can play forever and then the strategy 
is winning for her. Hence, the strategy is winning for game $\mathcal{G}_{A^M, A'}$ 
since $s_{init}\;\M\;s'_{init}$.

$\Leftarrow)$ Suppose that the verifier has a winning strategy 
from the initial state. Then, we define a masking simulation relation 
as $\M = \{(s,s') \mid \text{ \emph{V} has a winning stra-}$ \\
$\text{tegy for } (s, \#, s', R) \}$.
It is simple to see that it is a masking simulation. Furthermore, $s_{init} \; \M\; s'_{init}$, 
then $A \Masking A'$.	 \\

 \noindent
The proof is similar for the theorem that $A \WeakMasking A'$ iff the verifier has a winning strategy for the weak masking game graph $\mathcal{G}^W_{A^M, A'}$, but using 
$\mathcal{G}^W_{A^M, A'}$ instead of $\mathcal{G}_{A^M, A'}$ and by theorem~\ref{thm:weak_thm}.

\end{proof}\\

\noindent
\textbf{Proof of Theorem \ref{thm:mask_game_det}.}
  For any quantitative strong masking game $\mathcal{Q}_{A^M, A'}$ with payoff function $\FMask$,%
  \[\textstyle
  \Inf_{\pi_V \in \Pi_V} \; \Sup_{\pi_R \in \Pi_R} \FMask(out(\pi_R, \pi_V)) = \Sup_{\pi_R \in \Pi_R} \;  \Inf_{\pi_V \in \Pi_V} \FMask(out(\pi_R, \pi_V))\]

\begin{proof} In order to prove that the masking payoff function $\FMask$ is determined we have to prove that it is bounded and Borel measurable (Martin's theorem \cite{Martin98}). First, $\FMask$ is bounded by definition. Second, to see that $\FMask$ is Borel measurable note that $\FMask(\Omega) \subseteq [0,1]$, and then it is sufficient to prove that, for every rational $x$, $\FMask^{-1}((-\infty, x])$ is  Borel in the Cantor topology of infinite executions. 
Consider $\FMask^{-1}([-\infty,x])$ for an arbitrary $x$, this is the same as $\FMask^{-1}([0, \frac{1}{a}])$ for a given $a$. But, $\FMask^{-1}([0, \frac{1}{a}]) = \bigcup_{b \geq a} A_b$ where
$A_b = \bigcup_{i >0} A^i_b$ for $A^i_b = \{ \rho_0 \sigma_0 \rho_1 \sigma_1 \dots \mid \rho_i = s_{err} \wedge \sum^{i-1}_{j=0} \chi_{\mathcal{F}}(\sigma_j) =b\}$. Note that 
$A^i_b = \{ C_{\rho_0 \sigma_0 \dots \rho_i} \mid \sum^{i-1}_{j=0} \chi_{\mathcal{F}}(\sigma_j) =b\}$ where $C_{\rho_0 \sigma_0 \dots \rho_i}$ is the cone corresponding to initial segment 
$\rho_0 \sigma_0 \dots \rho_i$ which is Borel measurable, and so $A^i_b$, $A_b$ and $\FMask^{-1}((-\infty, x])$ are Borel measurable.
\end{proof} \\

\noindent
\textbf{Proof of Theorem \ref{thm:quant_game}.}
  Let $\mathcal{Q}_{A^M,A'}$ be a quantitative strong masking
  game.
  Then, $\val(\mathcal{Q}_{A^M,A'}) = \frac{1}{w}$, with
  $w = \min \{ i \mid \exists j : s_{init} \in U^j_i \}$, whenever
  $s_{init} \in U$, and $\val(\mathcal{Q}_{A^M,A'})=0$ otherwise, where
  sets $U^j_i$ and $U$ are defined in equation~(\ref{def:of:Uji}).\\

\begin{proof} 
	First, note that any play avoiding state $\ErrorSt$ has value $0$. By definition of the game, each transition performed by the Refuter  must be followed by a transition
selected by the Verifier. These transitions (the matches performed by the Verifier) have cost $(1,0)$ since the target of any of these transition is different from $\ErrorSt$. Because we have an infinite number of these matches, when state $\ErrorSt$ is not reached, the valuation
of these plays is $\lim_{n\rightarrow \infty} \frac{0}{1+ \sum^n_{i=0} v_i} = 0$.
	Otherwise, if $s_{init} \in U^j_i$ for some $j \geq 1$, we denote by $\Pi$  the set of Refuter's strategies satisfying the following:
	If $v \in U^j_i$ for $i,j>1$ and $post(v)  \cap U^{j-1}_{i} \neq \emptyset$, then 
	$\pi(v) = v'$, for some $v' \in post(v)  \cap U^{j-1}_{i}$.
	Note that $\Pi \neq \emptyset$, since any Refuter's node in a set $U^j_i$ has a successor belonging to $U^{j-1}_i$. Now, any play from $s_{init}$ following a strategy in $\Pi$
contains the occurrence of at most $i$ faults since the unique way of decreasing $i$ is by performing a masking after a fault, and $i \leq j$ always. That is, for any $\pi_V \in \Pi_{V}$ and 
$\pi_R \in \Pi$ we have that $\val(\pi_V, \pi_R) = \frac{1}{i}$. Thus, $\val(\mathcal{Q}_{A^M, A'}) \geq \frac{1}{i}$. Hence, $\val(\mathcal{Q}_{A^M, A'})\geq \frac{1}{w}$ for
$w = \min \{ i \mid \exists j : s_{init} \in U^j_i \} $. 

Now, note that for those nodes $s_i \notin U^i_j$ for every $i$ and $j$, the Verifier has strategies $\pi_V$ such that $\val(\pi_V, \pi_R)=0$ for any Refuter's strategy $\pi_R$. Then, for any Refuter's strategy $\pi_R \notin \Pi$ we have that $\inf_{\pi_R \in \Pi_R} \val(\pi_V, \pi_R) = 0$. That is, for any Refuter's strategy we have
$\inf_{\pi_V \in \Pi_V} \val(\pi_V, \pi_R) \leq \frac{1}{w}$ for $w=\min \{ i \mid \exists j : s_{init} \in U^j_i\}$. Therefore, $\sup_{\pi_R \in \Pi_R} \inf_{\pi_V \in \Pi_V} \val(\pi_V, \pi_R) \leq \frac{1}{w}$.
That is, $\val(\mathcal{Q}_{A^M,A'}) \leq \frac{1}{w}$, i.e., $\val(Q_{A^M,A'}) = \frac{1}{w}$.
\end{proof} \\

\noindent
\textbf{Proof of Theorem \ref{thm:triang_ineq}. }
  Let $A = \langle S, \Sigma, E, s_0 \rangle$, $A' = \langle S', \Sigma_{\mathcal{F'}}, E', s'_0 \rangle$,  and  $A'' = \langle S'', \Sigma_{\mathcal{F''}},E'', s''_0 \rangle$ be transition systems such that $\mathcal{F}' \subseteq \mathcal{F}''$.
  Then $\DeltaMask(A,A'') \leq \DeltaMask(A,A') + \DeltaMask(A', A'')$ and
  $\DeltaMask^W(A,A'') \leq \DeltaMask^W(A,A') + \DeltaMask^W(A', A'').$\\


\noindent
\begin{proof}
	Let us consider any node $(s, \#, s'', R)$ of the game $\mathcal{Q}_{A^M,A''}$ belonging to $U^j_i$ with $j \geq 2$. Note that $j$ cannot be the error state and so $j \neq 1$; moreover, after the movement of the Refuter we have at least one movement from the Verifier. In addition, for every nodes
$(s,\#, s', R)$ in $\mathcal{Q}_{A^M,A'}$ with $(s,\#, s', R) \in U_{i'}^{k'}$ and  $(s',\#,s'', R)$ of game $\mathcal{Q}_{{A'}^M,A''}$ with $(s',\#,s'', R) \in U^{k''}_{i''}$ it holds that
$\frac{1}{i} \leq \frac{1}{i'} + \frac{1}{i''}$. For the sake of convenience, when a node $s$ does not belong to any $U^k_i$, we assume $s \in U^{\infty}_{\infty}$. Then, we just define 
$\frac{1}{\infty}=0$.
The result follows from this fact and theorem \ref{thm:quant_game}. 
The proof is by induction on $i$.
\\ 
\noindent \emph{Base Case:} For $i=1$, we perform an induction on $j$. 
Let $j=2$ and suppose that $(s, \#, s'', R) \in U^2_1$. This means that we have a transition 
$(s, \#, s'', R) \xrightarrow{\sigma} (w, \sigma^t, w'', V)$, where $t \in \{1,2\}$, that cannot be matched by the Verifier. In case that $t=1$, then this play is a transition
$(s, \#, s'', R) \xrightarrow{\sigma^t} (w, \sigma^t, s'', V)$ from $A$. Now, let $(s, \#, s', R)$ and 
$(s', \#, s'', R)$ be a pair of nodes of $A$ and $A'$, respectively. 
By definition, we have a transition $(s, \#, s', R) \xrightarrow{\sigma} (w, \#, s', R)$ in $Q_{A^M,A'}$. 
In case that the Verifier cannot match this play in that game
we have that $(s, \#, s', R) \in U^2_1$. This finishes the proof since $1 \leq 1 + k''$, regardless of the value of $k''$. Otherwise, we have a play by the 
refuter  $(w, \#, s', R) \xrightarrow{\sigma^1} (w, \sigma^1, w', V)$ and we also have a transition $(s', \#, s'', R) \xrightarrow{\sigma} (w', \sigma, s'', V)$. But, this cannot be matched
by our initial assumption, that is, $(s', \#, s'', R) \in U^2_1$. This finalizes the base case for $j$.
For $t=2$,  the reasoning is similar using the transitions of $A''$. 
Now, for the inductive case of the second induction consider $j>2$ and $i=1$, that is, $(s, \#, s'', R) \in U^j_1$. This means that we have a transition
$(s, \#, s'', R) \xrightarrow{\sigma^t} (w, \sigma^t, w'', V)$ with $t \in \{1,2\}$. Consider now any pairs of states $(s, \#, s', R)$ in $\mathcal{Q}_{A,A'}$ and 
$(s', \#, s'', R)$ in $\mathcal{Q}_{A',A''}$. In case of $t=1$, then we 
have a transition $(s, \#, s'', R) \xrightarrow{\sigma^1} (w, \sigma^1, s'', V)$ where
$post((w, \sigma^1, s'', V)) \subseteq \bigcup^{k \leq j} U^j_1$. 
By definition of game $Q_{A^M, A'}$, we have a transition
$(s, \#, s', R) \xrightarrow{\sigma} (w, \sigma, s', V)$. In case that it cannot be matched, 
then the result follows. Otherwise, we have transitions
$(w, \sigma, s', V) \xrightarrow {\sigma} (w, \#, w', R)$ for $w' \in S'$. Therefore, 
there must be also a transition $(s', \#, s'', R) \xrightarrow{\sigma} (w', \sigma, s'', V)$. 
Similarly, in case that this cannot be matched we have $(s', \#, s'', R) \in U^2_1$ and the proof finishes. In other case, we have a collection of transitions
$(w', \sigma, s'', V) \xrightarrow{} (w', \#, w'', R)$ for $w'' \in S''$. Note 
that for any of  these pairs $(w, \#, w', R)$ and $(w', \#, w'', R)$, we have that
$(s, \#, s'', R) \xrightarrow{\sigma} (w,\sigma^1, s'', V) \xrightarrow{\sigma} (w, \#, w'', R)$. 
Then, $(w,\#,w'', R) \in U^{j-2}_1$ and by inductive hypothesis
for all of these pairs we have $(w, \#, w', R) \in U^{j'}_1$ and $(w', \#, w'', R) \in U^{j''}_1$,
Now, taking $k'$ as the maximum of all these $j'$ and $k''$ as the maximum of
all these $j''$, we obtain that $(w, \sigma, s', V)\in U^{k'+1}_1$ and $(w', \sigma, s'', V)\in U^{k''+1}_1$. This implies that $(s, \sigma, s', V)\in U^{k'+2}_1$ and $(s', \sigma, s'', V)\in U^{k''+2}_1$, which finishes the proof.
 \\ 
 \noindent \emph{Inductive Case:} For $i>1$ the proof is as follows. 
Assume that $(s, \#, s'', R) \in U^j_i$. Since $1<i < j$, we have a transition 
$(s, \#, s'', R) \xrightarrow{\sigma^t} (w, \sigma^t, w'', V)$. In case that 
$\sigma^t = F^2$ for some $F \in \mathcal{F}''$, then
we must have a transition $(s, F^2, w'', V) \xrightarrow{M} (s, \#, w'', R)$ and $(s, \#, w'', R) \in U^{j-2}_{i-1}$. On the other hand, in game $\mathcal{Q}_{A',A''}$
we must have a transition $(s', \#, s'', R) \xrightarrow{F} (s', F^2, w'', V)$ by definition.
In case that $F \in \mathcal{F}'$, then $F \in \Sigma'$. 
On the contrary, if it cannot be matched, then the result follows. 
Otherwise, we have a collection of transitions  
$(s', F^2, w'', V) \xrightarrow{F} (w', \#, w'', R)$. So, in game $\mathcal{Q}_{A,A'}$
we have at least an edge $(s, \#, s', R) \xrightarrow{F} (s, F^2, w', V)$. By the initial assumption, 
this can be masked, and then there is a transition $(s, F^2, w', V) \xrightarrow{M} (s, \#, w', R)$. 
By induction, we have $(s, \#, w', R) \in U^{j'}_{i'}$ and $(w', \#, w'', R) \in U^{j''}_{i''}$ such that
$\frac{1}{i-1} \leq \frac{1}{i'} + \frac{1}{i''}$. Note that $(s, F^2, w', V) \in U^{j'-1}_{i'-1}$, 
since we have a unique (masking) transition from $(s, F^2, V, w')$. Now, 
let us define $k'' = \max \{i'' \mid \text{for all states}~(s', \#, w'', V) \in U^{j''}_{i''} \}$. 
Then, $(s', F^2, s'', V) \in U^{k''}_{i''}$ and we have that $(s, \#, s', R) \in U^{j'}_{i'+1}$ and $(s', \#, s'', V) \in U^{k''}_{i''}$ with $\frac{1}{i} \leq \frac{1}{i'+1} + \frac{1}{i''}$ by definition of sets $U$'s. 
This finishes the proof for this case. 
In case that $\sigma^t \neq F^2$, the proof proceeds by induction on $j$ as 
the second induction in the base case. \\ 

 \noindent
The proof for $\DeltaMask^W$ is similar to the $\DeltaMask$ 
but using $\mathcal{Q}^W_{A^M,A'}$ instead of $\mathcal{Q}_{A^M,A'}$ and by theorem~\ref{thm:weak_thm}.

\end{proof}

\section{Models for Case Studies} \label{sec:case_studies}

In this section we provide models for some instances of the case studies presented in Section 4, these and models for other instances can be found on the tool repository.

\subsection{Memory Cell (3 bits)}

Here we have a basic model for a 3 bit redundancy memory cell, there is a single process Memory with actions for reading and writing a value. The process may fail by flipping one or more bits.

\begin{lstlisting}[
    basicstyle=\tiny,
]
Process Memory {
	w: BOOL; // the last value written, 
	r: BOOL; // the value we can read from the memory
	c0: BOOL; // the first bit
	c1: BOOL; // the second bit
	c2: BOOL; // the third bit
	Initial: w && c0 && c1 && c2 && r;
	Normative: (c0==c1) && (c1==c2) && (c0==c2) && w==r;
	[write] true -> w=!w, c0=!c0, c1=!c1, c2=!c2, r =!r;
	[read0] !r -> r = r;
	[read1] r -> r = r;
	[fail1] faulty true -> c0=!c0, r =(!c0&&c1)||(c1&&c2)||(!c0&&c2); 
	[fail2] faulty true -> c1=!c1, r =(c0&&!c1)||(!c1&&c2)||(c0&&c2); 
	[fail3] faulty true -> c2=!c2, r =(c0&&c1)||(c1&&!c2)||(c0&&!c2);    
}

Main(){
	m1: Memory;
	run m1();
}



\end{lstlisting}
\subsection{N-Modular Redundancy (3 modules)}

This is a model for 3-Modular Redundancy, there are three processes: Module,Voter and Environment. Modules can fail by flipping the input signal, The Voter outputs the majority value of the signals received, and the Environment can reset the input to 0 or 1.

\begin{lstlisting}[
    basicstyle=\tiny,
]
Global i0,i1,i2:BOOL; // inputs for each module

Process Module(out:BOOL) {
	Initial: !i0 && !i1 && !i2;
	Normative: true;
	[fail] faulty true -> out = !out;
}

Process Voter{
	Initial: !i0 && !i1 && !i2;
	Normative: true;
	[vote] (i0&&i1)||(i1&&i2)||(i0&&i2) -> i0 = i0; //if majority then skip
}

Process Environment{
	Initial: !i0 && !i1 && !i2;
	Normative: true;
	[input0] true -> i0 = false, i1 = false, i2 = false;
	[input1] true -> i0 = true, i1 = true, i2 = true;
}

Main(){
	m0: Module;
	m1: Module;
	m2: Module;
	v0: Voter;
	e0: Environment;
	run m0(i0);
	run m1(i1);
	run m2(i2);
	run v0();
	run e0();
}
\end{lstlisting} 
\subsection{Byzantine Agreement (4 generals)} 

This is a model for Byzantine Agreement with 4 generals, we differentiate the commander as a separate process Commander, and the other generals are instances of a Lieutenant process. In this case, there are two rounds of messages, the first one is an order from the commander to the lieutenants of attack or retreat, then comes the second round which involves a forward of the commander order from each lieutenant to all other lieutenants. Lieutenants may become traitors at any moment and send conflicted messages.

\begin{lstlisting}[
    basicstyle=\tiny,
]
Global g1g2A,g1g3A,g1g4A: BOOL; //Commander(g1) attack messages
Global g2g3A,g2g4A: BOOL; //Lieutenant1(g2) attack messages
Global g3g2A,g3g4A: BOOL; //Lieutenant2(g3) attack messages
Global g4g2A,g4g3A: BOOL; //Lieutenant3(g4) attack messages

Global g1g2R,g1g3R,g1g4R: BOOL; //Commander(g1) retreat messages
Global g2g3R,g2g4R: BOOL; //Lieutenant1(g2) retreat messages
Global g3g2R,g3g4R: BOOL; //Lieutenant2(g3) retreat messages
Global g4g2R,g4g3R: BOOL; //Lieutenant3(g4) retreat messages

Global A2,A3,A4: BOOL; //The Attack decision of each lieutenant
Global R2,R3,R4: BOOL; //The Retreat decision of each lieutenant

Process Commander{
	s0,s1: BOOL;
	Initial: s0 && !s1;
	Normative: true;
	[sA] s0 -> g1g2A = true, g1g3A = true, g1g4A = true, s0= false, s1=true;
	[sR] s0 -> g1g2R = true, g1g3R = true, g1g4R = true, s0= false, s1=true;
}


Process Lieutenant(attack: BOOL, retreat: BOOL, fw1A: BOOL, fw2A:BOOL, fw1R:BOOL, 
	fw2R:BOOL, a1:BOOL, a2:BOOL, r1:BOOL, r2:BOOL, dA:BOOL, dR:BOOL){
	// PARAMS: attack: attack order from commander, fw1A and fw2A: messages sent 
	// to other lieutenants, a1 and a2: messages received from other lieutenants, 
	// dA: decide to attack, the rest of params are similar but with retreat 
	s0,s1,s2, isBetrayer: BOOL;
	Initial: s0 && !s1 && !s2 && !isBetrayer;
	Normative: true;
	[fA] s0 && attack && !isBetrayer -> fw1A = true, fw2A= true, s0 = false, 
	s1 = true;
	[fR] s0 && retreat && !isBetrayer -> fw1R = true, fw2R = true, s0 = false, 
	s1 = true;
	[fA] s0 && attack && isBetrayer -> fw1R = true, fw2R= true, s0 = false, 
	s1 = true;
	[fR] s0 && retreat && isBetrayer -> fw1A = true, fw2A= true, s0 = false, 
	s1 = true;
	[Betray] faulty s0 && !isBetrayer -> isBetrayer = true; 
	[Attack] s1 && !isBetrayer && ((attack && a1)||(attack && a2)||(a1 && a2)) 
	-> s1 = false, s2 = true, dA = true;
	[Retreat] s1 && !isBetrayer && ((retreat && r1)||(retreat && r2)||(r1 && r2)) 
	-> s1 = false, s2 = true, dR = true;

}

Main(){
	g1:Commander;
	g2:Lieutenant;
	g3:Lieutenant;
	g4:Lieutenant;
	run g1();
	run g2(g1g2A,g1g2R,g2g3A,g2g4A,g2g3R,g2g4R,g3g2A,g4g2A,g3g2R,g4g2R,A2,R2);
	run g3(g1g3A,g1g3R,g3g2A,g3g4A,g3g2R,g3g4R,g2g3A,g4g3A,g2g3R,g4g3R,A3,R3);
	run g4(g1g4A,g1g4R,g4g3A,g4g2A,g4g3R,g4g2R,g2g4A,g3g4A,g2g4R,g3g4R,A4,R4);
}
\end{lstlisting} 
\subsection{Dining Philosophers (3 philosophers)}

Here we show a model for the Dining Philosophers in the case of 3 philosophers, 2 of which take the right fork first and the other takes the left one first, these are modeled as processes EvenPhil and OddPhil respectively. We incorporate a faulty behaviour on EvenPhil that makes it behave as OddPhil, i.e takes the left fork first.

\begin{lstlisting}[
    basicstyle=\tiny,
]
// !s0!s1 == thinking
// !s0s1 == hungry
// s0!s1 == eating
Global fork0,fork1,fork2:BOOL;

Process OddPhil(forkL:BOOL, forkR:BOOL){
	s0,s1 : BOOL;
	hasL, hasR : BOOL;
	Initial: !s0 && !s1 && !hasL && !hasR && forkR && forkL;
	Normative: !(hasR && !hasL);
	[hungry] !s0 && !s1 -> s1 = true;
	[getLeft] !s0 && s1 && forkL && !hasL && !hasR  -> forkL=false, hasL=true;
	[getRight] !s0 && s1 && hasL && forkR && !hasR -> forkR = false, hasR=true;
	[eating] !s0 && s1 && hasL && hasR -> s1 = false, s0 = true; 
	[thinking] s0 && !s1 -> s0 = false, forkL=true, forkR=true, hasR=false, 
	hasL=false;
}

Process EvenPhil(forkL:BOOL, forkR:BOOL){
	s0,s1 : BOOL;
	hasL, hasR : BOOL;
	Initial: !s0 && !s1 && !hasL && !hasR && forkR && forkL;
	Normative: !(hasL && !hasR); 
	[hungry] !s0 && !s1 -> s1 = true;
	[getRight] !s0 && s1 && forkR && !hasL && !hasR  -> forkR=false, hasR=true; 
	[getLeft] !s0 && s1 && hasR && forkL && !hasL -> forkL = false, hasL=true;
	[eating] !s0 && s1 && hasL && hasR -> s1 = false, s0 = true; 
	[thinking] s0 && !s1 -> s0 = false, forkL=true, forkR=true, hasR=false, 
	hasL=false;
	[getLeft] faulty !s0 && s1 && !hasR && forkL && !hasL -> forkL = false, 
	hasL=true;
}

Main(){
	phil1:OddPhil;
	phil2:EvenPhil;
	phil3:EvenPhil;
	run phil1(fork2, fork0);
	run phil2(fork0, fork1);
	run phil3(fork1, fork2);
}

\end{lstlisting} 
\subsection{Bounded Retransmission Protocol (1 chunk, 3 retransmissions)}

The BRP protocol sends a file in a number of chunks, but allows only a bounded number of retransmissions of each chunk, here we model an instance of this problem with 3 retransmissions on a single chunk. There are two processes, the Sender and the Receiver, the former has a set of internal actions that represent the case when a message is lost (i.e. a fault has occurred) and has to retrasmit it.

\begin{lstlisting}[
    basicstyle=\tiny,
]

// N==1 number of chunks
// MAX==3 number of retransmissions
Global fs,ls,bs: BOOL;
Global flagK,flagL: BOOL;

Process Sender {
    s0,s1,s2: BOOL; // state variables idle(000),nextframe(001),waitack(010),
    // retransmit(011),success(100),error(101),waitsync(110)
    srep0,srep1: BOOL; // srep variables bottom(00),nok(01),dk(10),ok(11)
    sab: BOOL;
    rt0,rt1: BOOL; // firstattempt(00),retransmission1(01),retransmission2(10),
    // retransmission3(11)
   	Initial: !s0 && !s1 && !s2 && !srep0 && !srep1 && !bs && !sab && !fs && !ls && 
   	!rt0 && !rt1 && !flagK && !flagL;
   	Normative: true;
	
	//idle
	[NewFile] !s0 && !s1 && !s2 -> s2 = true, srep0 = false, srep1 = false; 

	//next frame
	[sendChunk] !s0 && !s1 && s2 && !flagK -> s1 = true, s2 = false, fs = true, 
	ls = true, bs = sab, rt0 = false, rt1 = false, flagK = true;

	//wait ack
	[receiveAck] !s0 && s1 && !s2 && !flagK && flagL -> s0 = true, s1 = false, 
	sab = !sab, flagL = false; 
	[TOMsg] faulty !s0 && s1 && !s2 && flagK -> s2 = true, flagK = false;

	// retransmit
	[sendChunk] internal !s0 && s1 && s2 && !rt0 && !rt1 && !flagK -> s2 = false, 
	fs = true, ls = true, 
	bs = sab, rt1 = true, flagK = true; 
	[sendChunk] internal !s0 && s1 && s2 && !rt0 && rt1 && !flagK -> s2 = false, 
	fs = true, ls = true, 
	bs = sab, rt0 = true, rt1 = false, flagK = true;
	[sendChunk] internal !s0 && s1 && s2 && rt0 && !rt1 && !flagK -> s2 = false, 
	fs = true, ls = true, 
	bs = sab, rt1 = true, flagK = true; 
	[error] internal !s0 && s1 && s2 && rt0 && rt1  -> s0 = true, s1 = false, 
	srep0 = false, srep1 = true;
	[error] internal !s0 && s1 && s2 && rt0 && rt1  -> s0 = true, s1 = false, 
	srep0 = true, srep1 = false;

	// success
	[success] s0 && !s1 && !s2  -> s0 = false, srep0 = true, srep1 = true;

	// error
	[restart] s0 && !s1 && s2 -> s0 = false, s2 = false; 

}

Process Receiver {
	r0,r1,r2: BOOL; // newfile(000), fstsafe(001), framereceived(010), 
	// framereported(011), idle(100), finish(101)
	rrep0,rrep1,rrep2: BOOL; // bottom(000), fst(001), inc(010), ok(011), nok(100)
	fr,lr,br,rab,recv: BOOL;
	Initial: !r0 && !r1 && !r2 && !rrep0 && !rrep1 && !rrep2 && !fr && !lr && !br && 
	!rab && !recv && !fs && !ls && !bs && !flagK && !flagL;
	Normative: true;

	// new_file
	[receiveFirstChunk] !r0 && !r1 && !r2 && flagK && !flagL -> r2 = true, fr = fs, 
	lr = ls, br = bs, recv = true, flagK = false;

	// fst_safe_frame
	[e] !r0 && !r1 && r2 && !flagL -> r1 = true, r2 = false, rab = br;

	// frame_received
	[setIndication] !r0 && r1 && !r2 && rab==br && fr && !lr  && !flagL-> r2 = true,
	rrep0 = false, rrep1 = false, rrep2 = true;
	[setIndication] !r0 && r1 && !r2 && rab==br && !fr && !lr  && !flagL -> r2 = true, 
	rrep0 = false, rrep1 = true, rrep2 = false;
	[setIndication] !r0 && r1 && !r2 && rab==br && !fr && lr   && !flagL -> r2 = true, 
	rrep0 = false, rrep1 = true, rrep2 = true;
	[sendAck] !r0 && r1 && !r2 && !(rab==br)  && !flagL -> r0 = true, r1 = false, 
	flagL = true;  

	// frame_reported
	[sendAck] !r0 && r1 && r2 && !flagL && !lr -> r0 = true, r1 = false, r2 = false, 
	rab = !rab, flagL = true;
	[sendAck] !r0 && r1 && r2 && !flagL && lr -> r0 = true, r1 = false, r2 = true, 
	rab = !rab, flagL = true;

	// idle
	[receiveChunk] r0 && !r1 && !r2  && flagK && !flagL -> r0 = false, r1 = true, 
	fr = fs, lr = ls, br = bs, 
	recv = true,flagK = false;

	//finish
	[restart] r0 && !r1 && r2 -> r1 = false, r2 = false;

}

Main(){
	s: Sender;
	r: Receiver;
	run s();
	run r();
}

\end{lstlisting}

\end{document}